\def\etal{{\frenchspacing\it et al.}}

\def\beq#1{\begin{equation}\label{#1}}
\def\eeq{\end{equation}}
\def\beqa#1{\begin{eqnarray}\label{#1}}
\def\eeqa{\end{eqnarray}}

\def\la{\mathrel{\mathpalette\fun <}}

\def\fun#1#2{\lower3.6pt\vbox{\baselineskip0pt\lineskip.9pt
        \ialign{$\mathsurround=0pt#1\hfill##\hfil$\crcr#2\crcr\sim\crcr}}}
        
\def\xi{{{\bf x}^b}}

\documentclass[twocolumn,aps,showpacs,showkeys,nofootinbib]{revtex4}
\usepackage{epsfig}

\newcommand{\be}{\begin{equation}}
\newcommand{\ee}{\end{equation}}
\newcommand{\ba}{\begin{eqnarray}}
\newcommand{\ea}{\end{eqnarray}}

\begin{document}
\input{epsf.sty}

\title{Exploring uncertainties in dark energy constraints using current observational data with Planck 2015 distance priors}
\author{Yun~Wang$^{1,2}$\footnote{email: wang@ipac.caltech.edu}, 
Mi Dai$^{2}$}
\address{$^1$Infrared Processing and Analysis Center, California Institute of Technology,
770 South Wilson Avenue, Pasadena, CA 91125\\
$^{2}$Homer L. Dodge Department of Physics \& Astronomy, Univ. of Oklahoma, 440 W Brooks St., Norman, OK 73019}
                 
                 \today

\begin{abstract}

We present the distance priors that we have derived from the 2015 Planck data, and use these in combination with the
latest observational data from Type Ia Supernovae (SNe Ia) and galaxy clustering, to explore the systematic uncertainties
in dark energy constraints. We use the Joint Lightcurve Analysis (JLA) set of 740 SNe Ia, galaxy clustering measurements
of $H(z)s$ and $D_A(z)/s$ (where $s$ is the sound horizon at the drag epoch) from the Sloan Digital Sky Survey (SDSS)
at $z=0.35$ and $z=0.57$ (BOSS DR12).
We find that the combined dark energy constraints are insensitive to the assumptions made in the galaxy clustering 
measurements (whether they are for BAO only or marginalized over RSD), which indicates that as the analysis of
galaxy clustering data becomes more accurate and robust, the systematic uncertainties are reduced.
On the other hand, we find that flux-averaging SNe Ia at $z\geq 0.5$ significantly tightens the dark energy constraints,
and excludes a flat universe with a cosmological constant at 68\% confidence level, assuming a dark energy equation of state
linear in the cosmic scale factor. 
Flux-averaging has the most significant effect when we allow dark energy density function $X(z)$ to be a free functions given by the cubic
spline of its value at $z=0$, $\frac{1}{3}$, $\frac{2}{3}$, 1; the measured $X(z)$ deviates from a cosmological constant at more than 
95\% confidence level for $ 0.4 \la z \la 0.7$.
Since flux-averaging reduces the bias in the SN distance measurements, this may be an indication that we have arrived in the era
when the SN distance measurements are limited by systematic uncertainties.

\end{abstract}



\maketitle

\section{Introduction}

We continue to search for the unknown cause for the observed cosmic acceleration \cite{Riess98,Perl99},
a.k.a., dark energy.\footnote{For recent reviews, see 
\cite{Copeland06,Ruiz07,Ratra07,Frieman08,Caldwell09,Uzan09,Wang10,Li11,Weinberg12}.}
Current observational data offer tantalizing hints for deviations from a cosmological constant
in a simplistic combination of all observational data without critical analysis.
In order to arrive at robust constraints on dark energy, we must carefully examine all the data sets separately, and jointly.
One complication is the difficulty to detect and model unknown systematic uncertainties in the data used for the
analysis. 

In this paper, we explore the existence of unknown systematic uncertainties by critically analyzing the
latest observational data from Type Ia Supernovae (SNe Ia) and
galaxy clustering, with the help of distance priors from cosmic microwave background anisotropy (CMB) data.
We use the Joint Lightcurve Analysis (JLA) set of 740 SNe Ia, galaxy clustering measurements of $H(z)s$ and $D_A(z)/s$ (where $s$ is the
sound horizon at the drag epoch) from the Baryon Oscillation Spectroscopic Survey (BOSS) at $z=0.35$ and $z=0.57$, and the distance priors 
that we have derived from the 2015 Planck data.

We describe our method in Sec.II, present our results in Sec.III,
and conclude in Sec.IV.

\section{Method}
\label{sec:method}

We focus on exploring the unknown systematic uncertainties in the current SN Ia and galaxy clustering data using geometric constraints only, with distance priors from the 2015 Planck data
to help tighten parameter constraints. For a conservative and transparent approach, we marginalize over constraints on the growth rate of cosmic large scale structure (which 
are degenerate with the geometric constraints \citep{Wang08a,Simpson10}).

We now give the basic formulae that we will use later in the paper.
The comoving distance to an object at redshift $z$ is given by:
\ba
\label{eq:r(z)}
 & &r(z)=cH_0^{-1}\, |\Omega_k|^{-1/2} {\rm sinn}[|\Omega_k|^{1/2}\, \Gamma(z)],\\
 & &\Gamma(z)=\int_0^z\frac{dz'}{E(z')}, \hskip 1cm E(z)=H(z)/H_0 \nonumber
\ea
where ${\rm sinn}(x)=\sin(x)$, $x$, $\sinh(x)$ for 
$\Omega_k<0$, $\Omega_k=0$, and $\Omega_k>0$ respectively.
The Hubble parameter, $H(z)$, is given by
\ba
\label{eq:H(z)}
&&H^2(z)  \equiv  \left(\frac{\dot{a}}{a}\right)^2 \\
 &= &H_0^2 \left[ \Omega_m (1+z)^3 +\Omega_r (1+z)^4 +\Omega_k (1+z)^2 
+ \Omega_X X(z) \right],\nonumber
\ea
where $\Omega_m+\Omega_r+\Omega_k+\Omega_X=1$.
The dark energy density function $X(z) \equiv \rho_X(z)/\rho_X(0)$.
The $\Omega_r$ term,  with $\Omega_r=\Omega_m /(1+z_{eq}) \ll \Omega_m$ ($z_{eq}$ denotes the redshift at matter-rediation equality),
is usually omitted in dark energy studies at $z\ll 1000$, since dark energy should only be important at late times.
For comparison with the work of others and to provide a reference for future surveys, we 
consider a dark energy equation of state linear in the cosmic scale factor $a$ \cite{Chev01}:
\be
 w_X(a)=w_0+(1-a)w_a.
 \ee

In addition, we consider an alternative two parameter parametrization of $w_X(a)$, using $w_0$
and $w_{0.5} \equiv w_X(z=0.5)$:
\ba
w_X(a)&=&\left(\frac{\frac{2}{3}-a}{\frac{2}{3}-1}\right) w_0+ \left(\frac{a-1}{\frac{2}{3}-1}\right)w_{0.5}\nonumber\\
&=& 3w_{0.5}-2w_0+3(w_0-w_{0.5})a
\label{eq:w0wc}
\ea
Note that $a=\frac{2}{3}$ for $z=0.5$. It has been shown that $(w_0,w_{0.5})$ are significantly less correlated compared
to $(w_0, w_a)$ \cite{Wang08b}.

Finally, we consider a model-independent parametrization of $X(z)$, where $X(z)$ is a
free function of redshift given by the cubic spline of its value at $z=0$,
$\frac{1}{3}$, $\frac{2}{3}$, and $1$.  We assume that $X(z>1)=X(z=1)$.

\subsection{CMB data}
\label{sec:CMB}

We use CMB data in the condensed form of the CMB shift parameters (a.k.a., distance priors) \cite{WangPia07}:
\ba
R &\equiv &\sqrt{\Omega_m H_0^2} \,r(z_*)/c, \nonumber\\
l_a &\equiv &\pi r(z_*)/r_s(z_*).
\ea
These two parameters, $R$ and $l_a$, together with $\omega_b\equiv \Omega_b h^2$, provide an efficient summary
of CMB data as far as dark energy constraints go \cite{WangPia07,Li08}.

CMB data give us the comoving distance to the photon-decoupling surface 
$r(z_*)$, and the comoving sound horizon 
at photo-decoupling epoch $r_s(z_*)$ \cite{Page03}.
The comoving sound horizon at redshift $z$ is given by
\ba
\label{eq:rs}
r_s(z)  &= & \int_0^{t} \frac{c_s\, dt'}{a}
=cH_0^{-1}\int_{z}^{\infty} dz'\,
\frac{c_s}{E(z')}, \nonumber\\
 &= & cH_0^{-1} \int_0^{a} 
\frac{da'}{\sqrt{ 3(1+ \overline{R_b}\,a')\, {a'}^4 E^2(z')}},
\ea
where $a$ is the cosmic scale factor, $a =1/(1+z)$, and
$a^4 E^2(z)=\Omega_m (a+a_{\rm eq})+\Omega_k a^2 +\Omega_X X(z) a^4$,
with $a_{\rm eq}=\Omega_{\rm rad}/\Omega_m=1/(1+z_{\rm eq})$, and
$z_{\rm eq}=2.5\times 10^4 \Omega_m h^2 (T_{CMB}/2.7\,{\rm K})^{-4}$.
The sound speed is $c_s=1/\sqrt{3(1+\overline{R_b}\,a)}$,
with $\overline{R_b}\,a=3\rho_b/(4\rho_\gamma)$,
$\overline{R_b}=31500\Omega_bh^2(T_{CMB}/2.7\,{\rm K})^{-4}$.
We take $T_{CMB}=2.7255$.

The redshift to the photon-decoupling surface, $z_*$, is given by the 
fitting formula \cite{Hu96}:
\be
z_*=1048\, \left[1+ 0.00124 (\Omega_b h^2)^{-0.738}\right]\,
\left[1+g_1 (\Omega_m h^2)^{g_2} \right],
\ee
where
\ba
g_1 &= &\frac{0.0783\, (\Omega_b h^2)^{-0.238}}
{1+39.5\, (\Omega_b h^2)^{0.763}}\\
g_2 &= &\frac{0.560}{1+21.1\, (\Omega_b h^2)^{1.81}}
\ea
The redshift of the drag epoch $z_d$ is well approximated by \cite{EisenHu98}
\begin{equation}
z_d  =
 \frac{1291(\Omega_mh^2)^{0.251}}{1+0.659(\Omega_mh^2)^{0.828}}
\left[1+b_1(\Omega_bh^2)^{b2}\right],
\label{eq:zd}
\end{equation}
where
\begin{eqnarray}
  b_1 &= &0.313(\Omega_mh^2)^{-0.419}\left[1+0.607(\Omega_mh^2)^{0.674}\right],\\
  b_2 &= &0.238(\Omega_mh^2)^{0.223}.
\end{eqnarray}

Since the constraints on $(l_a, R, \omega_b, n_s)$ are {\it not} sensitive to the assumption about dark energy \cite{Wang12CM},
we are able to use the Planck archiv to obtain constraints on $(l_a, R, \omega_b, n_s)$ from the 2015 Planck data.
We use data from the Planck archive that include both temperature and polarization data, as well as CMB lensing.
As we have shown in earlier work \cite{WangPia07}, the one dimensional marginalized probability distributions of $(l_a, R, \omega_b, n_s)$
are well fitted by Gaussian distributions. For the Planck 2015 data,
$(l_a, R, \omega_b, n_s)$ are given by Gaussian distributions with the following means and standard deviations,
without assuming a flat Universe:
\ba
&&\langle l_a \rangle =           301.76, \sigma(l_a)=     0.093  \nonumber\\
&&\langle R \rangle =    1.7474, \sigma(R)=       0.0051 \nonumber\\
&& \langle \omega_b \rangle =    0.02228,  \sigma(\omega_b)=   0.00016 \nonumber\\
&& \langle n_s \rangle =    0.9659,  \sigma(n_s)=      0.0048
\label{eq:CMB_mean_planck}
\ea
with the normalized covariance matrix of $(l_a, R, \omega_b, n_s)$:
\be
\left(
\begin{array}{cccc}  
  1.0000  &   0.4529    & -0.3507 &   -0.3576\\  
  0.4529    &   1.0000  & -0.7000  & -0.7780\\  
 -0.3507    & -0.7000 &    1.0000 &      0.5296\\   
 -0.3576   &  -0.7780 &     0.5296  &     1.0000\\   
\end{array}
\right)
\label{eq:normcov_planck}
\ee

Assuming a flat Universe, the Planck 2015 data give $(l_a, R, \omega_b, n_s)$ well fit by Gaussian distributions with the following means and standard deviations:
\ba
&&\langle l_a \rangle =           301.77, \sigma(l_a)=     0.090  \nonumber\\
&&\langle R \rangle =    1.7482, \sigma(R)=       0.0048 \nonumber\\
&& \langle \omega_b \rangle =    0.02226,  \sigma(\omega_b)=   0.00016 \nonumber\\
&& \langle n_s \rangle =    0.9653,  \sigma(n_s)=      0.0048
\label{eq:CMB_mean_planck_flat}
\ea
with the normalized covariance matrix of $(l_a, R, \omega_b, n_s)$:
\be
\left(
\begin{array}{cccc}   
     1.0000   &   0.3996 &   -0.3181 &     -0.3004\\   
  0.3996 &       1.0000 &    -0.6891 &     -0.7677\\    
-0.3181 &    -0.6891 &       1.0000 &     0.5152 \\    
 -0.3004 &     -0.7677 &     0.5152 &      1.0000\\   
\end{array}
\right)
\label{eq:normcov_planck_flat}
\ee

We have included $n_s$ in our distance priors for completeness.
For the remainder of this paper,  we marginalize the CMB distance priors 
over $n_s$. This means dropping the 4th row and 4th column from
the normalized covariance matrix of $(l_a, R, \omega_b, n_s)$,
then obtain the covariance matrix for $(l_a, R, \omega_b)$ as follows:
\be
\mbox{Cov}_{CMB}(p_i,p_j)=\sigma(p_i)\, \sigma(p_j) \,\mbox{NormCov}_{CMB}(p_i,p_j),
\label{eq:CMB_cov}
\ee
where $i,j=1,2,3$. The rms variance $\sigma(p_i)$ and the normalized
covariance matrix $\mbox{NormCov}_{CMB}$ are given by
Eqs.(\ref{eq:CMB_mean_planck}) and (\ref{eq:normcov_planck})
without assuming a flat universe, and Eqs.(\ref{eq:CMB_mean_planck_flat}) and (\ref{eq:normcov_planck_flat})
for a flat universe.

We include the Planck distance priors by adding
the following term to the $\chi^2$ of a given model
with  $p_1=l_a$, $p_2=R$,and $p_3=\omega_b$:
\be
\label{eq:chi2CMB}
\chi^2_{CMB}=\Delta p_i \left[ \mbox{Cov}^{-1}_{CMB}(p_i,p_j)\right]
\Delta p_j,
\hskip .5cm
\Delta p_i= p_i - p_i^{data},
\ee
where $p_i^{data}$ are the mean from Eq.(\ref{eq:CMB_mean_planck}) (without assuming a flat universe) and
Eq.(\ref{eq:CMB_mean_planck_flat}) (assuming a flat universe),
and ${\rm Cov}^{-1}_{CMB}$ is the inverse of the covariance matrix of 
[$l_a, R, \omega_b]$ from Eq.(\ref{eq:CMB_cov}).
Note that $p_4=n_s$ should be added if the constraints on
$n_s$ are included in the galaxy clustering data.

\subsection{Analysis of SN Ia Data}

The distance modulus to a SN Ia is given by
\be
\label{eq:m-M}
\mu_0 \equiv m-M= 5 \log\left[\frac{d_L(z)}{\mathrm{Mpc}}\right]+25,
\ee
where $m$ and $M$ represent the apparent and absolute magnitude
of a SN. The luminosity distance $d_L(z)=(1+z)\, r(z)$, with the comoving
distance $r(z)$ given by Eq.(\ref{eq:r(z)}). 

We use the JLA set of 740 SNe Ia processed by Betoule et al. (2014) \cite{Betoule14}.
They give the apparent $B$ magnitude, $m_B$, and the covariance matrix for
$\Delta m \equiv m_B-m_{\rm mod}$, with \cite{Conley11}
\be
m_{\rm mod}=5 \log_{10}{\cal D}_L(z|\mbox{\bf s})
- \alpha X_1 +\beta {\cal C} + {\cal M},
\ee
where ${\cal D}_L(z|\mbox{\bf s})$ is the luminosity distance
multiplied by $H_0$
for a given set of cosmological parameters $\{ {\bf s} \}$,
$X_1$ is the stretch measure of the SN light curve shape, and
${\cal C}$ is the color measure for the SN.
${\cal M}$ is a nuisance parameter representing some combination
of the absolute magnitude of a fiducial SN Ia, $M$, and the 
Hubble constant $H_0$. ${\cal M}$ is assumed to be different for SNe Ia with
different host stellar mass: 
\ba
{\cal M}&=&M_1 \hskip 0.2in  \mbox{for host stellar mass} < 10^{10} M_{\odot}\nonumber\\
{\cal M}&=&M_2 \hskip 0.2in \mbox{otherwise}
\ea
Since the time dilation part of the observed luminosity distance depends 
on the total redshift $z_{\rm hel}$ (special relativistic plus cosmological),
we have \cite{Hui06}
\be
{\cal D}_L(z|\mbox{\bf s})\equiv c^{-1}H_0 (1+z_{\rm hel}) r(z|\mbox{\bf s}),
\ee
where $z$ and $z_{\rm hel}$ are the CMB restframe and heliocentric redshifts
of the SN. 

For a set of $N$ SNe with correlated errors, we have 
\be
\label{eq:chi2_SN}
\chi^2=\Delta \mbox{\bf m}^T \cdot \mbox{\bf C}^{-1} \cdot \Delta\mbox{\bf m}
\ee
where $\Delta \bf m$ is a vector with $N$ components, and
$\mbox{\bf C}$ is the $N\times N$ covariance matrix of the SNe Ia.

Note that $\Delta m$ is equivalent to $\Delta \mu_0$, since 
\be
\Delta m \equiv m_B-m_{\rm mod}
= \left[m_B+\alpha X_1 -\beta {\cal C}\right] - {\cal M}.
\ee

The total covariance matrix is \cite{Conley11}
\be
\mbox{\bf C}=\mbox{\bf D}_{\rm stat}+\mbox{\bf C}_{\rm stat}
+\mbox{\bf C}_{\rm sys},
\ee
with the diagonal part of the statistical uncertainty given by \cite{Conley11,Betoule14}
\ba
\mbox{\bf D}_{\rm stat,ii}&=&\sigma^2_{m_B,i}+\sigma^2_{\rm int}
+ \sigma^2_{\rm lensing}  + \left[\frac{5}{z_i \ln 10}\right]^2 \sigma^2_{z,i}\nonumber\\
&&   +\alpha^2 \sigma^2_{X_1,i}+\beta^2 \sigma^2_{{\cal C},i} 
+ 2 \alpha C_{m_B X_1,i} - 2 \beta C_{m_B {\cal C},i}\nonumber\\
&&  -2\alpha\beta C_{X_1 {\cal C},i},
\ea
where $C_{m_B X_1,i}$, $C_{m_B {\cal C},i}$, and $C_{X_1 {\cal C},i}$
are the covariances between $m_B$, $X_1$, and ${\cal C}$ for the $i$-th SN. 
Note the Betoule et al. (2014) included host galaxy correction in 
$\mbox{\bf C}_{\rm stat}+\mbox{\bf C}_{\rm sys}$ (see Eq.(11) of \cite{Betoule14}).

The statistical and systematic covariance matrices, 
$\mbox{\bf C}_{\rm stat}$ and $\mbox{\bf C}_{\rm sys}$,
are generally not diagonal \cite{Conley11}, and are given in the
form:
\be
\mbox{\bf C}_{\rm stat}+\mbox{\bf C}_{\rm sys}
=V_0+\alpha^2 V_a + \beta^2 V_b + 2 \alpha V_{0a}
-2 \beta V_{0b} - 2 \alpha\beta V_{ab}.
\ee
where $V_0$, $V_{a}$, $V_{b}$, $V_{0a}$, $V_{0b}$, and
$V_{ab}$ are matrices given by Betoule et al.  at the link
http://supernovae.in2p3.fr/sdss$_{-}$snls$_{-}$jla/ReadMe.html,
$\mbox{\bf C}_{\rm stat}$ includes the uncertainty in
the SN model. $\mbox{\bf C}_{\rm sys}$ includes the
uncertainty in the zero point. Note that  $\mbox{\bf C}_{\rm stat}$
and $\mbox{\bf C}_{\rm sys}$ do not depend on ${\cal M}$, 
since the relative distance moduli are independent of the value 
of ${\cal M}$ \cite{Conley11}.

We refer the reader to Conley et al. (2011) \cite{Conley11}
and Betoule et al. (2014) \cite{Betoule14}
for detailed discussions of the origins of the statistical
and systematic errors. 

In order to explore the existence of unknown systematic effects, we apply flux-averaging to the JLA SNe Ia
at $z\geq 0.5$. Flux-averaging was proposed to reduce the systematic bias in distance measurement due to weak lensing magnification
of SNe Ia \cite{Wang00,WangPia04,Wang05}; it has the additional benefit of reducing the bias in distance
estimate due to other, possibly unknown systematic effects \cite{WangTegmark05}.
This is because flux-averaging effectively reduces a 
global systematic bias into a local bias with a much smaller amplitude,
which in turn results in a reduced impact on global parameter constraints.
Since weak lensing does not have a significant effect on SN Ia data (see, e.g.,
\cite{Sarkar07}),  any systematic biases in the current SN Ia data are likely dominated by 
other, presently unknown sources.

Here we apply flux-averaging in the minimal approach of flux-averaging the SNe Ia in each redshift bin at higher $z$,
and then use the usual ``magnitude statistics'' (instead of ``flux statistics''  \cite{Wang00,WangPia04,Wang05})
in computing $\chi^2$,
since the JLA SNe Ia have measurement and modeling errors that have been effectively Gaussianized in
magnitudes.

For $\chi^2$ statistics using MCMC or a grid of parameters, 
here are the steps in flux-averaging \cite{Wang12CM} in application to the JLA SNe Ia:

(1) Convert the distance modulus of SNe Ia into 
``fluxes'',
\be
\label{eq:flux}
F(z_l) \equiv 10^{-(\mu_0^{\rm data}(z_l)-25)/2.5} =  
\left( \frac{d_L^{\rm data}(z_l)} {\mbox{Mpc}} \right)^{-2}.
\ee

(2) For a given set of cosmological parameters $\{ {\bf s} \}$,
obtain ``absolute luminosities'', \{${\cal L}(z_l)$\}, by
removing the redshift dependence of the ``fluxes'', i.e.,
\be
\label{eq:lum}
{\cal L}(z_l) \equiv d_L^2(z_l |{\bf s})\,F(z_l).
\ee

(3) Flux-average the ``absolute luminosities'' \{${\cal L}^i_l$\} 
in each redshift bin $i$ to obtain $\left\{\overline{\cal L}^i\right\}$:
\be 
 \overline{\cal L}^i = \frac{1}{N_i}
 \sum_{l=1}^{N_i} {\cal L}^i_l(z^{(i)}_l),
 \hskip 1cm
 \overline{z_i} = \frac{1}{N_i}
 \sum_{l=1}^{N_i} z^{(i)}_l. 
\ee

(4) Place $\overline{\cal L}^i$ at the mean redshift $\overline{z}_i$ of
the $i$-th redshift bin, now the binned flux is
\be
\overline{F}(\overline{z}_i) = \overline{\cal L}^i /
d_L^2(\overline{z}_i|\mbox{\bf s}).
\ee
with the corresponding flux-averaged distance modulus:
\be
\overline\mu^{data}(\overline{z}_i) =-2.5\log_{10}\overline{F}(\overline{z}_i)+25.
\ee

(5) Compute the covariance matrix of $\overline{\mu}(\overline{z}_i)$
and $\overline{\mu}(\overline{z}_j)$:
\ba
&& \mbox{Cov}\left[\overline{\mu}(\overline{z}_i),\overline{\mu}(\overline{z}_j)\right] \\
&=&\frac{1}{N_i N_j \overline{\cal L}^i \overline{\cal L}^j}
 \cdot \nonumber\\
&& \sum_{l=1}^{N_i} \sum_{m=1}^{N_j} {\cal L}(z_l^{(i)})
{\cal L}(z_m^{(j)}) \langle \Delta \mu_0^{\rm data}(z_l^{(i)})\Delta 
\mu_0^{\rm data}(z_m^{(j)})
\rangle \nonumber 
\ea
where $\langle \Delta \mu_0^{\rm data}(z_l^{(i)})\Delta \mu_0^{\rm data}(z_m^{(j)})\rangle $
is the covariance of the measured distance moduli of the $l$-th SN Ia
in the $i$-th redshift bin, and the $m$-th SN Ia in the $j$-th
redshift bin. ${\cal L}(z)$ is defined by Eqs.(\ref{eq:flux}) and (\ref{eq:lum}).

(6) For the flux-averaged data, $\left\{\overline{\mu}(\overline{z}_i)\right\}$, 
compute
\be
\label{eq:chi2_SN_fluxavg}
\chi^2 = \sum_{ij} \Delta\overline{\mu}(\overline{z}_i) \,
\mbox{Cov}^{-1}\left[\overline{\mu}(\overline{z}_i),\overline{\mu}(\overline{z}_j)
\right] \,\Delta\overline{\mu}(\overline{z}_j)
\ee
where
\be
\Delta\overline{\mu}(\overline{z}_i) \equiv 
\overline{\mu}(\overline{z}_i) - \mu^p(\overline{z}_i|\mbox{\bf s}),
\ee
and
\be 
\overline\mu^p(\overline{z}_i) =-2.5\log_{10} F^p(\overline{z}_i)+25.
\ee
with $F^p(\overline{z}_i|\mbox{\bf s})=
\left( d_L(z|\mbox{\bf s}) /\mbox{Mpc} \right)^{-2}$.

For the sample of SNe we use in this study, we  
flux-averaged the SNe with $dz=0.04$.

\subsection{Galaxy Clustering Data}
\label{sec:GC}

For GC data, we use the measurements of $x_h(z)=H(z)r_s(z_d)/c$ and $x_d(z)=D_A(z)/r_s(z_d)$, 
where $H(z)$ is the Hubble parameter, $D_A(z)$ is the angular diameter distance, 
and $r_s(z_d)$ is the sound horizon at the drag epoch. It has been shown that $x_h(z)$ and $x_d(z)$ 
are more tightly constrained by data, and less sensitive to modeling assumptions, compared to
$H(z)$ and $D_A(z)$ \citep{CW12}.
We use the $x_h(z)$ and $x_d(z)$ measurements from the two-dimensional 
power spectrum measured at z=0.32 and z=0.57 from BOSS DR12 galaxies \citep{Gil-Marin2016a,Gil-Marin2016b}.
Converting the results in \citep{Gil-Marin2016a,Gil-Marin2016b} to the same definitions used in this paper, 
we find that for BAO only \citep{Gil-Marin2016a}
\ba
x_h (0.32) \equiv H(0.32)r_s(z_d)/c&=&0.0397	\pm  0.0021 \nonumber \\
x_d (0.32) \equiv D_A(0.32)/r_s(z_d)&=& 6.49	\pm  0.16 \nonumber\\
r_{hd}(0.32) &=&0.41\\
& & \nonumber\\
x_h (0.57) \equiv H(0.57)r_s(z_d)/c&=&0.0498	\pm  0.0013 \nonumber \\
x_d (0.57) \equiv D_A(0.57)/r_s(z_d)&=& 9.18	\pm  0.13 \nonumber\\
r_{hd}(0.57) &=&0.47
\label{eq:BAO}
\ea
For BAO measurements marginalized over RSD \citep{Gil-Marin2016b}, we find
\ba
x_h (0.32) \equiv H(0.32)r_s(z_d)/c&=&0.0391	\pm  0.0019 \nonumber \\
x_d (0.32) \equiv D_A(0.32)/r_s(z_d)&=& 6.185	\pm  0.185 \nonumber\\
r_{hd}(0.32) &=&0.5\\
& & \nonumber\\
x_h (0.57) \equiv H(0.57)r_s(z_d)/c&=&0.0476	\pm  0.0015 \nonumber \\
x_d (0.57) \equiv D_A(0.57)/r_s(z_d)&=& 9.18	\pm  0.15 \nonumber\\
r_{hd}(0.57) &=&0.53
\label{eq:BAO/RSD}
\ea

Galaxy clustering data are included in our analysis by adding $\chi^2_{GC}=\chi^2_{GC1}+\chi^2_{GC2}$,
with $z_{GC1}=0.35$ and $z_{GC2}=0.57$, to the $\chi^2$ of a given model.
Note that
\be
\label{eq:chi2bao}
\chi^2_{GCi}=\Delta p_i \left[ {\rm C}^{-1}_{GC}(p_i,p_j)\right]
\Delta p_j,
\hskip .5cm
\Delta p_i= p_i - p_i^{data},
\ee
where $p_1=H(z_{GCi})r_s(z_d)/c$ and $p_2=D_A(z_{GCi})/r_s(z_d)$,
with $i=1,2$.

\section{Results}

We perform a MCMC likelihood analysis \cite{Lewis02} to obtain 
${\cal O}$($10^6$) samples for each set of results presented in 
this paper. We assume flat priors for all the parameters, and allow ranges 
of the parameters wide enough such that further increasing the allowed 
ranges has no impact on the results.
We constrain dark energy and cosmological parameters ($w_0$, $w_a$, $\Omega_m, \Omega_k, h, \omega_b$), 
where $\omega_b\equiv \Omega_b h^2$. In addition, we marginalize over
the SN Ia nuisance parameters $\{\alpha, \beta, M_1, M_2\}$.

\subsection{Constrains on $w_0$ and $w_a$}

\begin{figure} 
\psfig{file=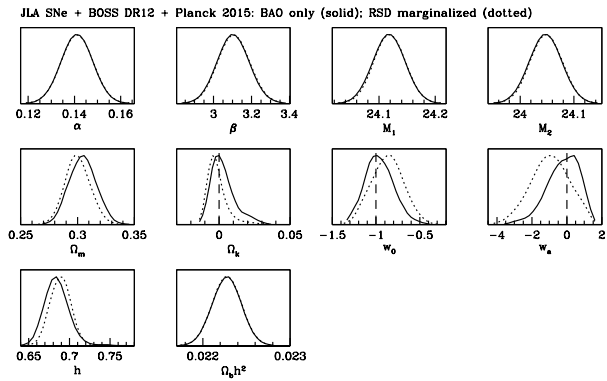,width=3.5in}\\
\vspace{-1.5in}
\caption{\footnotesize%
Marginalized probability distributions of parameters from JLA SNe, galaxy clustering data at $z=0.35$ and 
$z=0.57$, and Planck 2015 distance priors. The solid and dotted curves correspond to using $H(z)$ and $D_A(z)$ measurements 
from BAO only measurements, and those from RSD marginalized measurements.
}
\label{fig:baorsd:w0wa}
\end{figure}

\begin{figure} 
\psfig{file=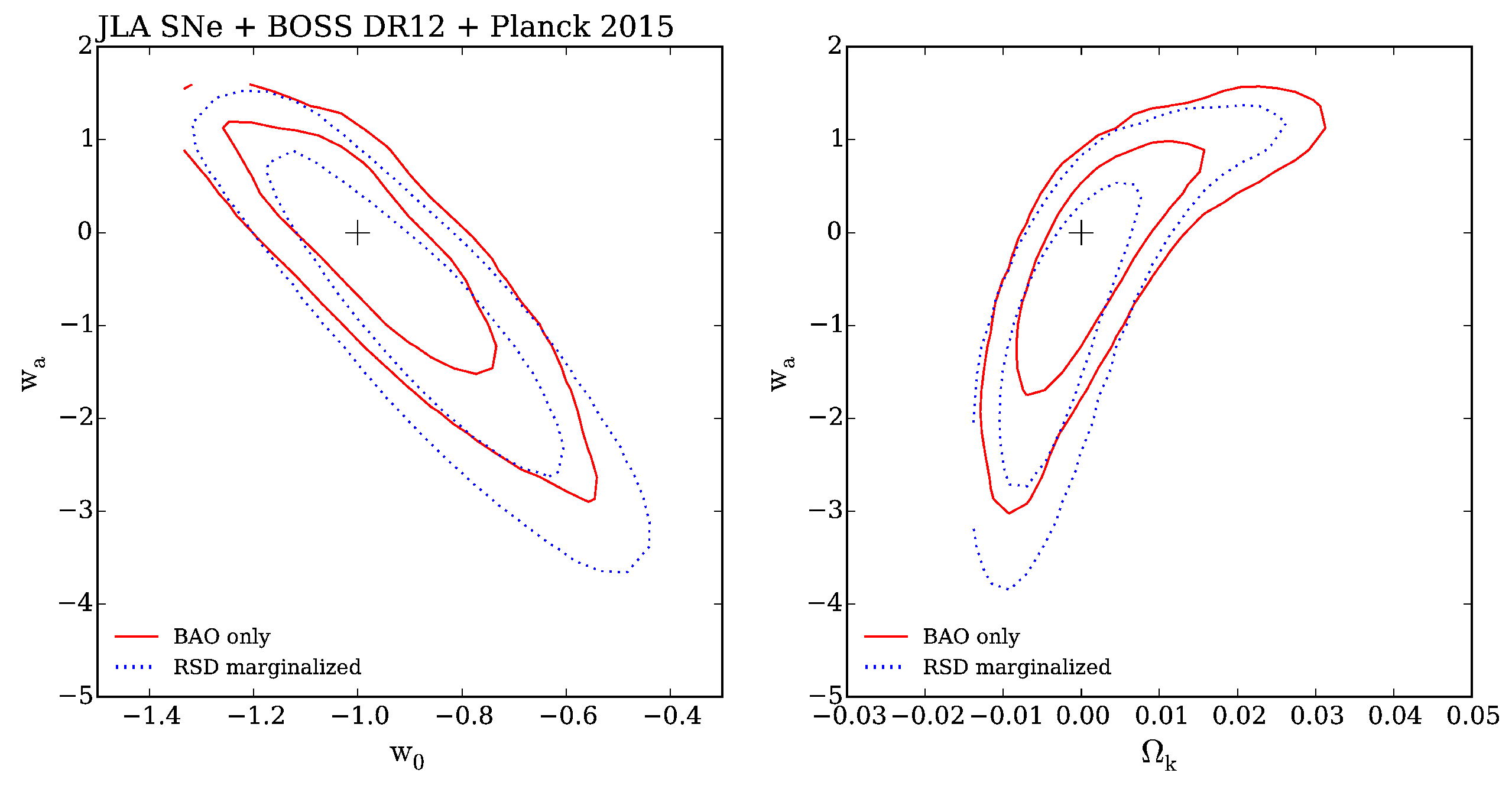,width=3.5in}\\
\caption{\footnotesize%
The joint 68\% and 95\% confidence contours for ($w_a$, $w_0$) and ($w_a$, $\Omega_k$) corresponding to Fig.\ref{fig:baorsd:w0wa},
with the same line types.
}
\label{fig:baorsd:w0wa_2D}
\end{figure}

Fig.\ref{fig:baorsd:w0wa} shows the marginalized probability distributions of parameters from JLA SNe, galaxy clustering data at $z=0.35$ and 
$z=0.57$ \cite{Gil-Marin2016a,Gil-Marin2016b}, and Planck 2015 distance priors presented in this paper (see Sec.IIA). 
The solid and dotted curves correspond to using $H(z)$ and $D_A(z)$ measurements 
from BAO only measurements, and those from RSD marginalized measurements.
Fig.\ref{fig:baorsd:w0wa_2D} shows the joint 68\% and 95\% confidence contours for ($w_a$, $w_0$) and ($w_a$, $\Omega_k$) corresponding to Fig.\ref{fig:baorsd:w0wa},
with the same line types. The combined dark energy constraints seem insensitive to the assumptions made in the analysis of galaxy clustering data.

\begin{figure} 
\psfig{file=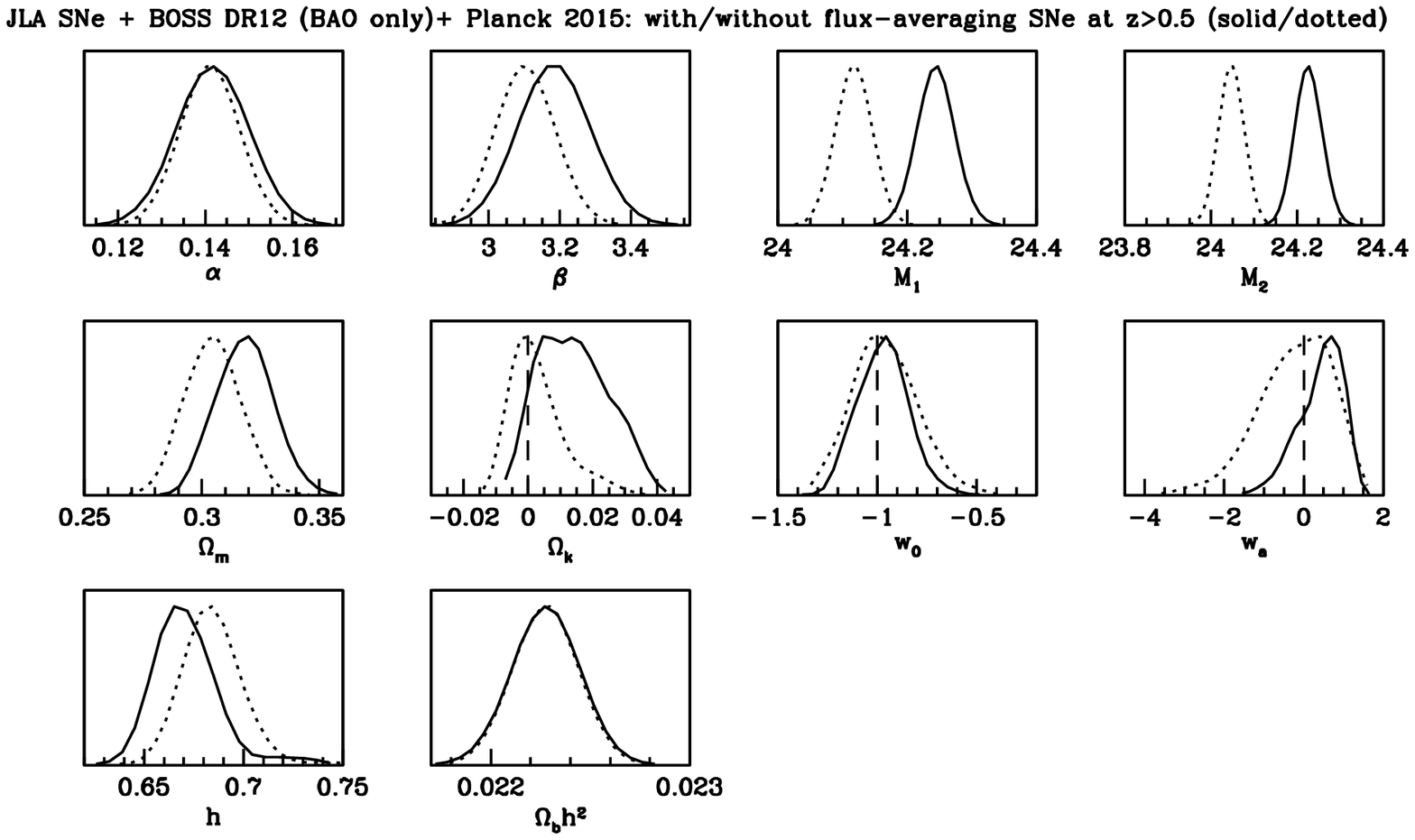,width=3.5in}\\
\vspace{-1.5in}
\caption{\footnotesize%
Marginalized probability distributions of parameters from JLA SNe, galaxy clustering data at $z=0.35$ and 
$z=0.57$ (BAO only), and Planck 2015 distance priors. The solid and dotted curves correspond to using SNe Ia
with and without flux-averaging.
}
\label{fig:fa:w0wa}
\end{figure}

\begin{figure} 
\psfig{file=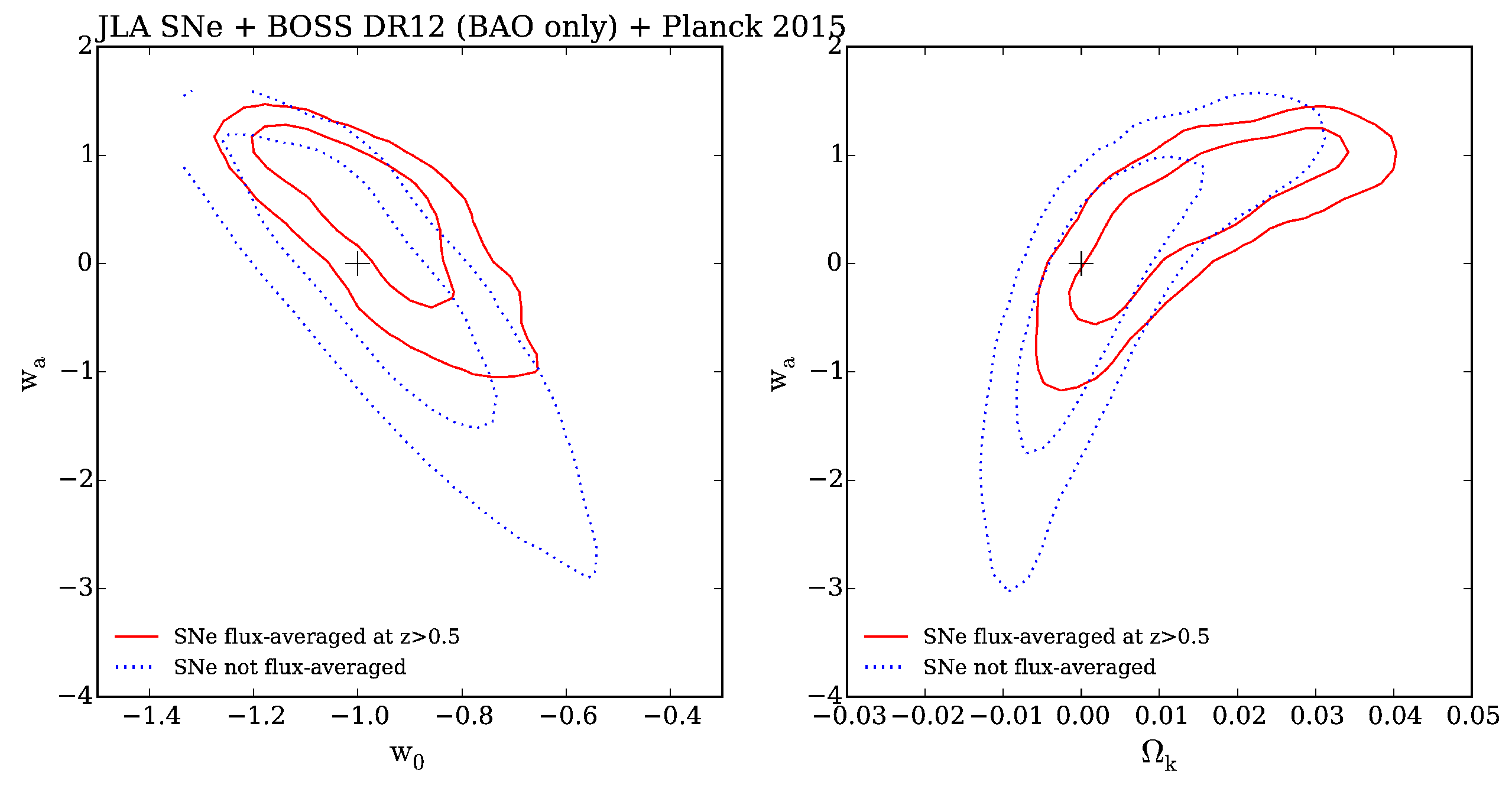,width=3.5in}\\
\caption{\footnotesize%
The joint 68\% and 95\% confidence contours for ($w_a$, $w_0$) and ($w_a$, $\Omega_k$) corresponding to Fig.\ref{fig:fa:w0wa}.
The solid and dotted curves correspond to using SNe Ia
with and without flux-averaging.
}
\label{fig:fa:w0wa_2D}
\end{figure}

Fig.\ref{fig:fa:w0wa} shows the impact of flux-averaging SNe Ia on the marginalized probability distributions of parameters from the combination
of the same data sets as in Fig.\ref{fig:baorsd:w0wa}. The solid and dotted curves correspond to using SNe Ia
with and without flux-averaging.
Fig.\ref{fig:fa:w0wa_2D} shows the joint 68\% and 95\% confidence contours for ($w_a$, $w_0$) and ($w_a$, $\Omega_k$) corresponding to Fig.\ref{fig:fa:w0wa},
with the same line types.
Clearly, flux-averaging significantly tightens the dark energy constraints. This may be due to the reduction in the bias of distance measurements from flux-averaging,
which increases the concordance of the data, resulting in tighter constraints.

\subsection{Constraints on $w_0$ and $w_{0.5}$}

\begin{figure} 
\psfig{file=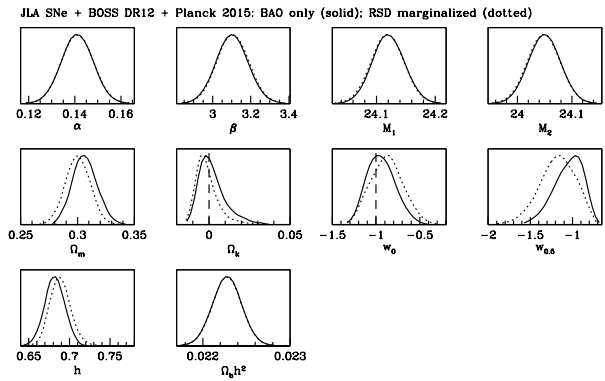,width=3.5in}\\
\vspace{-1.5in}
\caption{\footnotesize%
Marginalized probability distributions of parameters from JLA SNe, galaxy clustering data at $z=0.35$ and 
$z=0.57$, and Planck 2015 distance priors. The solid and dotted curves correspond to using $H(z)$ and $D_A(z)$ measurements 
from BAO only measurements, and those from RSD marginalized measurements.
}
\label{fig:baorsd:w0wc}
\end{figure}

\begin{figure} 
\psfig{file=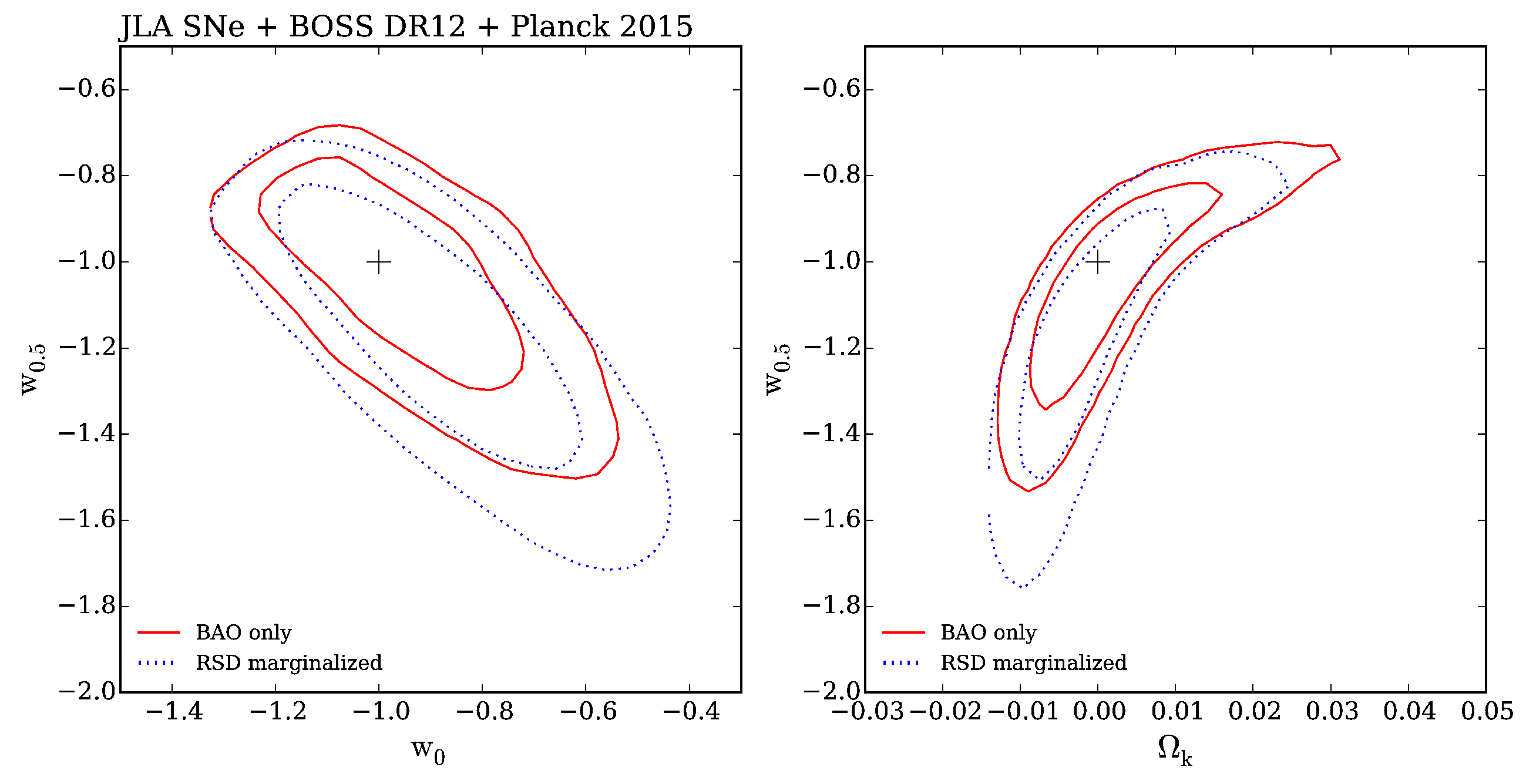,width=3.5in}\\
\caption{\footnotesize%
The joint 68\% and 95\% confidence contours for ($w_{0.5}$, $w_0$) and ($w_{0.5}$, $\Omega_k$) corresponding to Fig.\ref{fig:baorsd:w0wc}.
The solid and dotted curves correspond to using $H(z)$ and $D_A(z)$ measurements 
from BAO only measurements, and those from RSD marginalized measurements.
}
\label{fig:baorsd:w0wc_2D}
\end{figure}

\begin{figure} 
\psfig{file=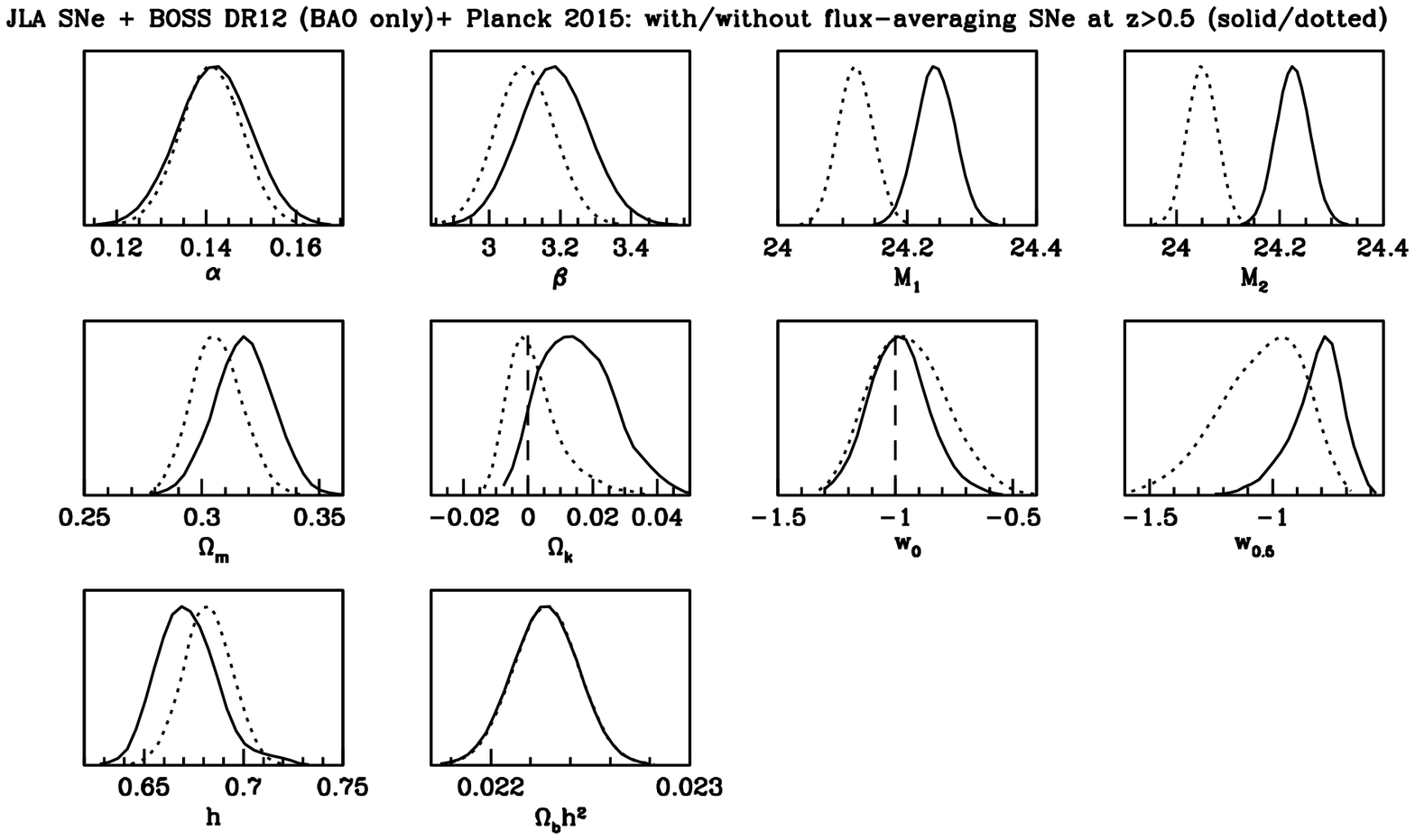,width=3.5in}\\
\vspace{-1.5in}
\caption{\footnotesize%
Marginalized probability distributions of parameters from JLA SNe, galaxy clustering data at $z=0.35$ and 
$z=0.57$ (BAO only), and Planck 2015 distance priors. The solid and dotted curves correspond to using SNe Ia
with and without flux-averaging.
}
\label{fig:fa:w0wc}
\end{figure}

\begin{figure} 
\psfig{file=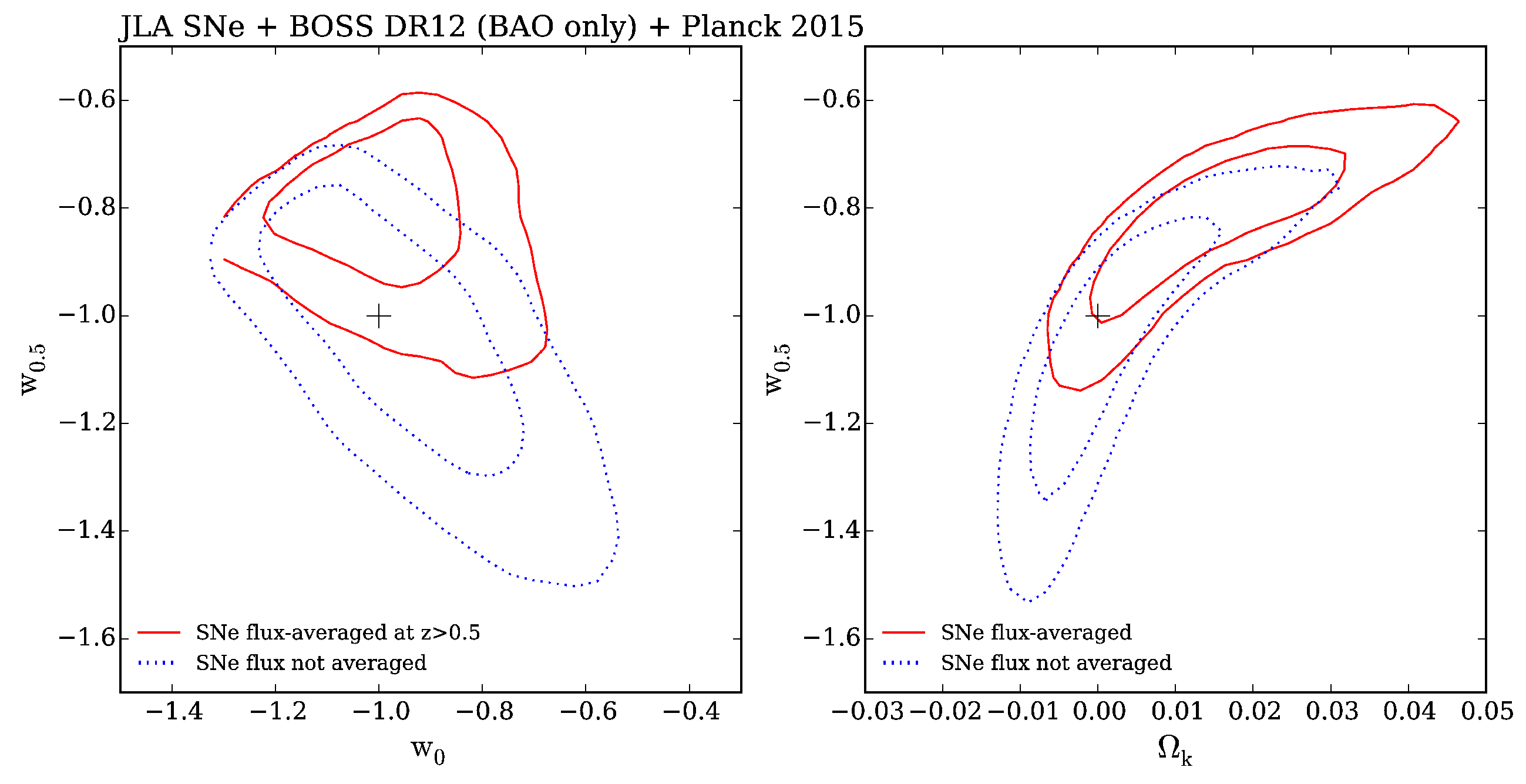,width=3.5in}\\
\caption{\footnotesize%
The joint 68\% and 95\% confidence contours for ($w_{0.5}$, $w_0$) and ($w_{0.5}$, $\Omega_k$) corresponding to Fig.\ref{fig:fa:w0wc}.
The solid and dotted curves correspond to using SNe Ia
with and without flux-averaging.
}
\label{fig:fa:w0wc_2D}
\end{figure}

Figs.\ref{fig:baorsd:w0wc}-\ref{fig:fa:w0wc_2D} are similar to Figs.\ref{fig:baorsd:w0wa}-\ref{fig:fa:w0wa_2D}, but for parametrizing the linear dark energy density
uisng $w_0$ and $w_{0.5}$ (see Eq.[\ref{eq:w0wc}]), instead of the usual $w_0$ and $w_a$.
Fig.\ref{fig:baorsd:w0wc} shows the impact of the galaxy clustering analysis technique on the marginalized probability distributions of parameters from JLA SNe, galaxy clustering data at $z=0.35$ and 
$z=0.57$ \cite{Gil-Marin2016a,Gil-Marin2016b}, and Planck 2015 distance priors presented in this paper (see Sec.IIA). 
The solid and dotted curves correspond to using $H(z)$ and $D_A(z)$ measurements 
from BAO only measurements, and those from RSD marginalized measurements.
Fig.\ref{fig:baorsd:w0wc_2D} shows the joint 68\% and 95\% confidence contours for ($w_{0.5}$, $w_0$) and ($w_{0.5}$, $\Omega_k$) corresponding to Fig.\ref{fig:baorsd:w0wc},
with the same line types. Again, the assumptions made in the analysis of galaxy clustering data have little impact on the combined dark energy constraints.

Fig.\ref{fig:fa:w0wc} shows the impact of flux-averaging SNe Ia on the marginalized probability distributions of parameters from the combination of the same data sets as in Fig.\ref{fig:baorsd:w0wc}. 
The solid and dotted curves correspond to using SNe Ia
with and without flux-averaging.
Fig.\ref{fig:fa:w0wc_2D} shows the joint 68\% and 95\% confidence contours for ($w_{0.5}$, $w_0$) and ($w_{0.5}$, $\Omega_k$) corresponding to Fig.\ref{fig:fa:w0wc},
with the same line types.
Flux-averaging of SNe makes an even more dramatic difference in the joint constraints on ($w_0$, $w_{0.5}$), compared to that of ($w_0, w_a)$..

\subsection{Constraints on dark energy density function}

We now examine the dark energy constraints when we allow the dark energy density function, $X(z)\equiv \rho_X(z)/\rho_X(z=0)$, to be a free function, 
given by the cubic spline of its value at $z=0, 1/3, 2/3, 1$, and assuming that $X(z>1)=X(z=1)$.

\begin{figure} 
\psfig{file=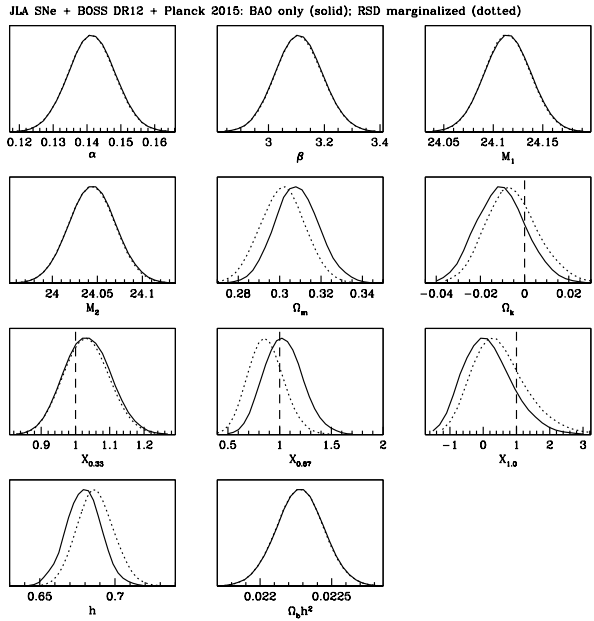,width=3.5in}\\
\caption{\footnotesize%
Marginalized probability distributions of parameters from JLA SNe, galaxy clustering data at $z=0.35$ and 
$z=0.57$, and Planck 2015 distance priors. The solid and dotted curves correspond to using $H(z)$ and $D_A(z)$ measurements 
from BAO only measurements, and those from RSD marginalized measurements.
}
\label{fig:baorsd:X3}
\end{figure}

\begin{figure} 
\psfig{file=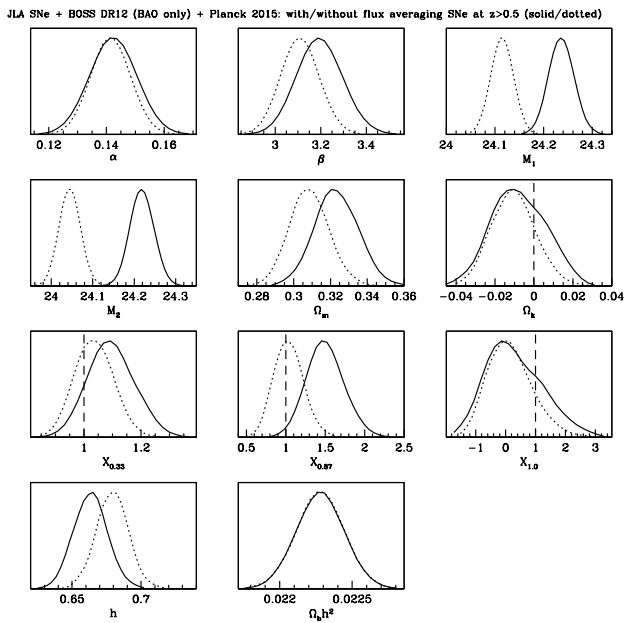,width=3.5in}\\
\caption{\footnotesize%
Marginalized probability distributions of parameters from JLA SNe, galaxy clustering data at $z=0.35$ and 
$z=0.57$ (BAO only), and Planck 2015 distance priors. The solid and dotted curves correspond to using SNe Ia
with and without flux-averaging.
}
\label{fig:fa:X3}
\end{figure}

Fig.\ref{fig:baorsd:X3} shows the impact of the galaxy clustering analysis technique on the marginalized probability distributions of parameters from JLA SNe, galaxy clustering data at $z=0.35$ and 
$z=0.57$ \cite{Gil-Marin2016a,Gil-Marin2016b}, and Planck 2015 distance priors presented in this paper (see Sec.IIA). 
The solid and dotted curves correspond to using $H(z)$ and $D_A(z)$ measurements 
from BAO only measurements, and those from RSD marginalized measurements. 
Fig.\ref{fig:fa:X3} shows the impact of flux-averaging the SNe Ia on the marginalized probability distributions of parameters from the same combination of data sets. The solid and dotted curves correspond to using SNe Ia
with and without flux-averaging.
Again, we find that the assumptions made in the galaxy clustering data analysis have little impact on the combined dark energy constraints,
while flux-averaging of SNe Ia has a significant impact on these constraints.

\begin{figure} 
\psfig{file=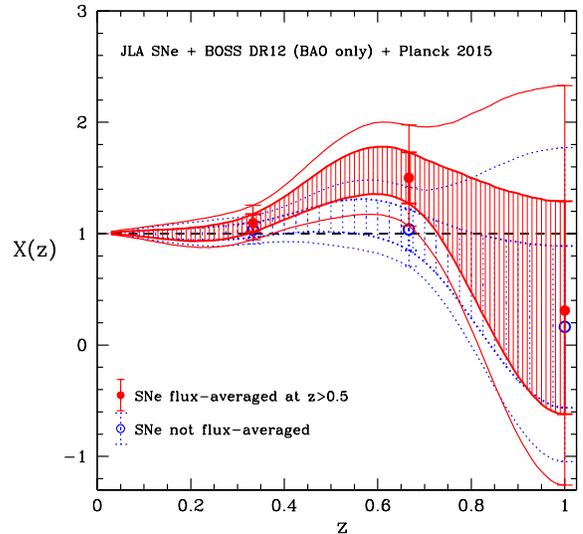,width=3.5in}\\
\vspace{-0.2in}
\caption{\footnotesize%
The dark energy density function $X(z)=\rho_X(z)/\rho_X(0)$ measured from JLA SNe, galaxy clustering data at $z=0.35$ and 
$z=0.57$ \cite{Gil-Marin2016a,Gil-Marin2016b}, and Planck 2015 distance priors presented in this paper (see Sec.IIA). The shaded regions
indicate the 68\% confidence region, while the outer envelope indicates the 95\% confidence level.
The densely shaded and sparsely shaded regions correspond to using SNe Ia with and without flux-averaging, respectively.
}
\label{fig:fa:Xz}
\end{figure}

Fig.\ref{fig:fa:Xz} shows the dark energy density function $X(z)=\rho_X(z)/\rho_X(0)$ measured from JLA SNe, galaxy clustering data at $z=0.35$ and 
$z=0.57$ \cite{Gil-Marin2016a,Gil-Marin2016b}, and Planck 2015 distance priors presented in this paper (see Sec.IIA). The shaded regions
indicate the 68\% confidence region, while the outer envelope indicates the 95\% confidence level.
The densely shaded and sparsely shaded regions correspond to using SNe Ia with and without flux-averaging, respectively.
Flux-averaging has the most significant effect here --- the measured $X(z)$ deviates from $X(z)=1$ ($w=-1)$ at more than 95\% confidence level
for $ 0.4 \la z \la 0.7$.

\section{Discussion and Summary}

We have explored the existence of unknown systematic uncertainties in the current SN Ia and galaxy clustering data,
with the help of the latest CMB distance priors. We use the JLA set of 740 SNe Ia from Betoule et a. (2014) \cite{Betoule14},
and the measurements of $H(z)$ and $D_A(z)$ at $z=0.35$ and $z=0.57$ from BOSS DR12 data by Gil-Marin et al.
2016 \citep{Gil-Marin2016a,Gil-Marin2016b}.
We have derived the CMB distance priors from Planck 2015 data, in the
form of the mean values and covariance matrix of $\{l_a, R,\Omega_b h^2, n_s\}$,
which give an efficient summary of Planck data in the context of dark energy
constraints (see Eqs.(\ref{eq:CMB_mean_planck}-\ref{eq:normcov_planck_flat})). 

It is remarkable that the Planck distance priors that we have derived from the 2015 Planck data have uncertainties that are within 10\% of the
forecasted errors for Planck by Mukherjee et al. (2008) \cite{Pia08}. This indicates that Planck has achieved its forecasted precision in
cosmological constraints. We note that Huang, Wang, \& Wang \cite{Huang15} independently derived similar but slightly different
constraints from Planck 2015 data at approximately the same time.

We find that the combined dark energy constraints are insensitive to the assumptions made in the galaxy clustering 
measurements (whether they are for BAO only \cite{Gil-Marin2016a} or marginalized over RSD \cite{Gil-Marin2016b}),
independent of the dark energy parametrization used (see Fig.\ref{fig:baorsd:w0wa}, Fig.\ref{fig:baorsd:w0wa_2D}, 
Fig.\ref{fig:baorsd:w0wc}, Fig.\ref{fig:baorsd:w0wc_2D}, Fig.\ref{fig:baorsd:X3}).
We note that the published BAO only constraints in \cite{Gil-Marin2016a} differ from those in the earlier arXiv version, and 
are closer to the RSD marginalized constraints in \cite{Gil-Marin2016b}. This is reassuring, as it indicates that as the analysis of
galaxy clustering data becomes more accurate and robust, the systematic uncertainties are reduced.

On the other hand, we find that flux-averaging SNe Ia at $z\geq 0.5$ significantly tightens the dark energy constraints,
 and excludes $w=-1$ at greater than 68\% confidence level (see Fig.\ref{fig:fa:w0wa}, 
Fig.\ref{fig:fa:w0wa_2D}, Fig.\ref{fig:fa:w0wc}, Fig.\ref{fig:fa:w0wc_2D}, Fig.\ref{fig:fa:X3}, Fig.\ref{fig:fa:Xz}).
Flux-averaging has the most significant effect when we allow dark energy density function $X(z)=\rho_X(z)/\rho_X(0)$ to be a free functions given by the cubic
spline of its value at $z=0$, $\frac{1}{3}$, $\frac{2}{3}$, 1; the measured $X(z)$ deviates from $X(z)=1$ ($w=-1)$ at more than 
95\% confidence level for $ 0.4 \la z \la 0.7$ (see Fig.\ref{fig:fa:Xz}).
This is somewhat surprising, since for SN data with redshift-dependent systematic biases that are negligible compared to statistical errors, 
flux-averaging of SNe should give somewhat less stringent constraints on dark energy \cite{Wang00}.
Since flux-averaging reduces the bias in the SN distance measurements \cite{WangTegmark05}, this may be an indication
that we have arrived in the era when the SN distance measurements are limited by systematic uncertainties.

Identifying and correctly modeling systematic effects will be key in illuminating the nature of dark energy.
Future dark energy surveys from space \cite{jedi,Cimatti09,euclid,wfirst} will be designed to
minimize systematic uncertainties. We can expect dramatic progress in the next decade in our quest to shed light on dark energy.

\bigskip

{\bf Acknowledgements}
We are grateful to Rick Kessler and Ranga Chary for helpful discussions, and to Alex Merson for providing python scripts for making 2D contour plots.
We acknowledge the use of Planck data archiv and CosmoMC.
     


\begin{thebibliography}{}

\bibitem[Riess et al.~(1998)]{Riess98}
Riess, A. G, {\etal}, 1998, Astron. J., 116, 1009

\bibitem[Perlmutter et al.~(1999)]{Perl99} 
Perlmutter, S. {\etal}, 1999, ApJ, 517, 565


\bibitem{Copeland06}
Copeland, E.~J., Sami, M., Tsujikawa, S., IJMPD, 15 (2006), 1753

\bibitem{Ruiz07}
Ruiz-Lapuente, P., Class. Quantum. Grav., 24 (2007), 91 

\bibitem{Ratra07}
Ratra, B., Vogeley, M.~S., arXiv:0706.1565 (2007)

\bibitem{Frieman08}
Frieman, J., Turner, M., Huterer, D., ARAA, 46, 385 (2008)

\bibitem{Caldwell09}
Caldwell, R. R., \& Kamionkowski, M., arXiv:0903.0866

\bibitem{Uzan09}
Uzan, J.-P., arXiv:0908.2243

\bibitem{Wang10}
Wang, Y., {\it Dark Energy}, Wiley-VCH (2010)

\bibitem{Li11}
Li, M., et al., 2011, arXiv1103.5870	

\bibitem{Weinberg12}
Weinberg, D. H.; et al., Physics Reports, in press, arXiv:1201.2434


\bibitem[Wang(2008a)]{Wang08a}
Wang, Y., Journal of Cosmology and Astroparticle Physics, 05, 021 (2008).

\bibitem[Simpson \& Peacock(2010)]{Simpson10}
Simpson, F., \& Peacock, J.A. 2010, Phys Rev D, 81, 043512  


\bibitem[Chev01(2001)]{Chev01}
Chevallier, M., \& Polarski, D. 2001, Int. J. Mod. Phys. D10,
213

\bibitem{Wang08b}
Wang, Y., 2008b, Phys. Rev. D 77, 123525 
Figure of Merit for Dark Energy Constraints from Current Observational Data



\bibitem[Wang \& Mukherjee(2007)]{WangPia07}
Wang, Y., \& Mukherjee, P., PRD, 76, 103533 (2007)

\bibitem{Li08}
Li, H., et al., ApJ, 683, L1 (2008)

\bibitem[Page(2003)]{Page03}
Page, L., et al. 2003, ApJS, 148, 233 


\bibitem[Hu \& Sugiyama(1996)]{Hu96}
Hu, W., \& Sugiyama, N. 1996, ApJ, 471, 542

\bibitem[Eisenstein \& Hu(1998)]{EisenHu98}
Eisenstein, D. \& Hu, W. 1998, ApJ, 496, 605

\bibitem[Wang, Chuang, \& Mukherjee(2012)]{Wang12CM}
Wang, Y.; Chuang, C.-H.; \& Mukherjee, P., Phys. Rev. D 85, 023517 (2012)

\bibitem{CW12}
Chuang, C.-H.; and Wang, Y., MNRAS, 426, 226 (2012)

\bibitem[Gil-Marin, et al.(2016a)]{Gil-Marin2016a}	
Gil-Marin, H., et al., 2016, MNRAS, 460, 4210

\bibitem[Gil-Marin, et al.(2016b)]{Gil-Marin2016b}	
Gil-Marin, H., et al., 2016, MNRAS, 460, 4188


\bibitem{Betoule14}
Betoule, M., et al., A \& A, 568, A22 (2014)

\bibitem{Conley11}
Conley, A., et al., 2011, Astrophys.J.Suppl., 192, 1

\bibitem{Hui06}
Hui, L.; \& Green, P.~B., PRD, 73, 123526 (2006)

\bibitem[Wang(2000)]{Wang00}
Wang, Y., ApJ 536, 531 (2000)

\bibitem[Wang \& Mukherjee(2004)]{WangPia04}
Wang, Y., \& Mukherjee, P. 2004, ApJ, 606, 654

\bibitem[Wang(2005)]{Wang05}
Wang, Y., JCAP, 03, 005 (2005)

\bibitem[Wang \& Tegmark(2005)]{WangTegmark05}
Wang, Y., \& Tegmark, M. , Phys. Rev. D 71, 103513 (2005)

\bibitem{Sarkar07}
Sarkar, D., et al., 2008, ApJ., 678, 1

\bibitem{Anderson14}
Anderson, L., et al., MNRAS, 441, 24 (2014)


\bibitem[Lewis02(2002)]{Lewis02}
Lewis, A., \& Bridle, S. 2002, PRD, 66, 103511

\bibitem{Pia08}
Mukherjee, P.;  Kunz, M.; Parkinson, D.; Wang, Y., 
Phys.Rev.D, 78, 083529 (2008)

\bibitem{Huang15}
Huang, Q.~G., Wang, K.,  and Wang, S., JCAP, 12, 022 (2015)


\bibitem{jedi}
Crotts, A. et al., 2005, astro-ph/0507043

\bibitem{Cimatti09}
Cimatti, A., et al., Experimental Astronomy, 23, 39 (2009)

\bibitem{euclid}
Laureijs, R.; et al., 2011, arXiv1110.3193

\bibitem{wfirst}
Spergel, D., et al., eprint arXiv:1503.03757



\end{thebibliography}
\end{document}